\documentstyle[12pt]{article}
\input BoxedEPS
\SetRokickiEPSFSpecial  
\HideDisplacementBoxes
\newcommand{\inputepsf}[1]{\BoxedEPSF{#1}}
\begin{document}
\thispagestyle{empty}
\noindent\hspace*{\fill}  OHSTPY-HEP-TH-94-014\\
\hspace*{\fill}  DOE/ER/01545-632\\
\begin{center}\begin{Large}\begin{bf}
            Spontaneous symmetry breaking \\
     of (1+1)-dimensional $\bf \phi^4$ theory \\
         in light-front field theory (III)\\
\end{bf}\end{Large}\vspace{.75cm}
\vspace{0.5cm}
              Stephen S. Pinsky and Brett van de Sande\\[10pt]
\vspace{0.1cm}
                    {\em Department of Physics\\
                   The Ohio State University\\
                      174 West 18th Avenue\\
                       Columbus, OH~~43210}\\
 \vspace{0.1cm}
\vspace{0.5cm}
              John R. Hiller\\[10pt]
\vspace{0.1cm}
                    {\em Department of Physics\\
                   University of Minnesota -- Duluth\\
                   Duluth, MN~~55812}\\
\vspace{0.1cm}
\end{center}
\vspace{1cm}\baselineskip=35pt
\begin{abstract} \noindent
We investigate (1+1)-dimensional $\phi^4$ field theory in the
symmetric and broken phases using discrete light-front quantization.
We calculate the
perturbative solution of the zero-mode constraint equation for both
the symmetric and broken phases and show that
standard renormalization of the theory yields finite results. We study
the perturbative zero-mode contribution to two diagrams and show
that the light-front formulation gives the same result as the equal-time
formulation.  In the broken
phase of the theory, we obtain the nonperturbative solutions of the constraint
equation and confirm our previous speculation that the critical coupling is
logarithmically divergent. We discuss the renormalization of this divergence
but are not able to find a satisfactory nonperturbative technique. Finally we
investigate properties that are insensitive to this divergence, calculate
the critical exponent of the theory, and find agreement with mean field theory
as expected.

\end{abstract}
\newpage\baselineskip=18pt
\section{Introduction}
\setcounter{equation}{0}
This paper is the third in a series of studies investigating one particular
long range phenomenon, spontaneous symmetry breaking, in a
particularly simple theory, $\phi^4$  theory in (1+1)-dimensions.
The $Z_2$ symmetry $\phi \to -\phi$ of the theory is spontaneously broken
for a range of mass and coupling.
At a critical coupling, this theory
develops spontaneously new interactions that do not respect the
$Z_2$ symmetry. In light-front field theory,
this symmetry breaking is produced by the zero mode of the
field. It can be shown that this zero mode is not an independent degree of
freedom in the theory but depends on all the other degrees of freedom through
a nonlinear operator constraint equation. In previous papers, which we
will refer to as I~\cite{BPV} and
II~\cite{PV}, we have shown how to solve this operator-valued
constraint equation
numerically and, in lowest order, analytically. We have investigated
spontaneous
symmetry breaking in low-order numerical calculations and discussed many of the
properties of a spontaneously broken theory, including the critical coupling
and
tunnelling. Here we conclude this work with a perturbative analysis and
higher order numerical calculations.

We provide a review of notation and basic ideas in the next section.
In Sec.~III, we discuss the perturbative behavior
of the theory. We  investigate the role of the zero mode in the
perturbative regime and
show that the usual mass subtraction renders the theory finite to all
orders even when the zero-mode interactions are included. We
also investigate the role of the zero-mode
interactions in two loop diagrams. In Sec.~IV
we study the nonperturbative structure of the theory.  We
also look at the behavior of the matrix elements of the
zero modes and calculate the critical coupling using a longitudinal
momentum cutoff. Finally, we
discuss the possible renormalization schemes
for the broken phase of the theory.  A summary of our findings
is given in Sec.~V.
\section{Review} \label{Review}
\setcounter{equation}{0}
The details of the Dirac-Bergmann prescription and its
application to the system considered in this paper are discussed
elsewhere in the literature~\cite{BPV,maskawa,heinzl,wittman}.
In this section, we will summarize those results,
introduce our notation, and review some of the results of I and II.
We define light-front coordinates
$x^\pm = \left(x^0 \pm x^1\right)/\sqrt{2}$.
For a classical field the $\left(\phi^4\right)_{1+1}$
Lagrangian is
\begin{equation}
{\cal L} = \partial_+\phi\partial_-\phi - {{\mu^2}\over 2}
\phi^2 - {\lambda
\over 4!} \phi^4\;.
\end{equation}
We put the system in a box of length $d$ and impose
periodic boundary conditions.  That is, we use
``discretized light-cone quantization'' (DLCQ) to
regulate the system~\cite{PauliBrodsky,PW}.
Following the Dirac-Bergmann prescription, we can identify
first-class constraints which define the conjugate momenta
and a secondary constraint
which determines the ``zero mode'' in terms of the other
modes of the theory.
The latter result can also be obtained by integrating the
equations of motion
in position space or differentiating the Hamiltonian with
respect to the zero mode~\cite{BPV}.

Quantizing, we define creation and annihilation operators
$a_k^\dagger$ and $a_k = a_{-k}^\dagger$, along with the zero-mode
operator $a_0$, by
\begin{equation}
\phi\!\left(x\right) = {a_0\over{\sqrt{4 \pi}}}+ {1\over{\sqrt{4 \pi}}}
\sum_{n \neq 0} {a_n(x^+)\over \sqrt{|n|}}\,
{ e}^{ik_n^+ x^-}\; ,
\end{equation}
where $k_n^+ =2\pi n/d$ and summations run over all integers
unless otherwise noted.
These operators satisfy the usual commutation relations
\begin{equation}
\left[a_k,a_l\right] =0\;,
\quad\left[a_k^\dagger,a_l^\dagger\right] =0\;,\quad
\left[a_k,a_l^\dagger\right] =\delta_{k, l}\; ,\quad k,l > 0\; .
\end{equation}
The light-front momentum operator $P^+$ is then given by
\begin{equation}
 P^+ = \frac{2 \pi}{d} \sum_{k>0} k a_k^\dagger a_k \; \; .
\end{equation}

The quantum Hamiltonian that we use~\cite{BPV} is obtained
from the light-front time evolution operator $P^-$
by a rescaling
\begin{eqnarray}
\label{rescale}
\lefteqn{H ={{96 \pi^2}\over{\lambda d}} P^- } \nonumber\\
&=&\frac{g}{2} a_0^2 + \frac{a_0^4}{4}  + g \Sigma_2+ 6 \Sigma_4
  \nonumber\\
& &+\frac{1}{4}\sum_{n\neq 0} {1\over{|n|}} \left( a_0^2 a_n a_{-n}
+ a_n a_{-n}
a_0^2 + a_n a_0^2 a_{-n} \right. \nonumber\\
& & \quad \left. +  a_n a_0 a_{-n} a_0 + a_0 a_n a_0 a_{-n} +
a_0 a_n a_{-n} a_0 - 3 a_0^2\right) \nonumber\\
& & +\frac{1}{4}\,\sum_{k, l, m \neq 0}\, {\delta_{k+l+m, 0}
\over{\sqrt{\left|k l m \right|}}}
{\left(a_0 a_k a_l a_m + a_k a_0 a_l a_m + a_k a_l a_0 a_m +
a_k a_l a_m a_0 \right)}   \nonumber \\
& &   -C  \; .
\end{eqnarray}
where $g= 24 \pi \mu^2/\lambda$ and $\Sigma_n$ is given by
\begin{equation}
\Sigma_n = \frac{1}{n!} \sum_{i_1, i_2, \ldots, i_n \neq 0}
\delta_{i_1+i_2+\cdots+i_n,0} \; \frac{: a_{i_1} a_{i_2} \cdots
a_{i_n}:}{
\sqrt{\left| i_1 i_2 \cdots i_n\right|}} \; .
\end{equation}
General arguments suggest that the Hamiltonian
should be symmetric ordered~\cite{benderpinsky}.
However, it is not clear how one should treat the
zero mode since it is not a dynamical field.
As an {\em ansatz} we treat $a_0$ as an ordinary
field operator when symmetric ordering the Hamiltonian.
We have removed tadpoles by
normal ordering the third and fourth terms and subtracting
\begin{equation}
{3\over{4}}\; a_0^2\sum_{n\neq 0} {1\over{|n|}} \; .
\end{equation}
In addition, we have subtracted a constant $C$ so that the
vacuum expectation value of $H$ is zero.
Note that this renormalization prescription is equivalent
to a conventional mass renormalization and
does not introduce any new operators (aside from the constant)
into the Hamiltonian.

The constraint equation for the zero mode can be determined by taking a
derivative of $P^-$ with respect to $a_0$.
\begin{equation}
0 = g a_0 + a_0^3 + 2 a_0 \Sigma_2 + 2 \Sigma_2 a_0 + \sum_{n > 0} {1\over n}
\left(  a_n^\dagger a_0 a_n + a_n a_0 a_n^\dagger - a_0 \right)+
6 \Sigma_3 \; . \label{constraint}
\end{equation}
Using the constraint equation, we can rewrite $H$ as:
\begin{eqnarray}
H &=& g \Sigma_2 + 6 \Sigma_4 -{a_0^4\over 4}
+\frac{1}{4}\sum_{n\neq 0} \frac{1}{|n|}
\left( a_n a_0^2 a_{-n} - a_0 a_n a_{-n} a_0\right) \nonumber \\
 & & +\frac{1}{4} \sum_{k, l, m \neq 0} \frac{\delta_{k+l+m, 0}}
{\sqrt{|k l m |}}
\left( a_k a_0 a_l a_m + a_k a_l a_0 a_m\right)  -  C \; .
 \label{hamiltonian}
\end{eqnarray}

In light-front field theory  the vacuum of the full theory is the perturbative
Fock space  vacuum.  In I and II, using $\left(\phi^4\right)_{1+1}$ as an
example, we have shown that the zero mode, which satisfies an operator-valued
constraint equation, produces the long range physics of the theory, including
spontaneous breaking of the  $Z_2$ symmetry. We have found that the constraint
equation can be solved  using a Tamm--Dancoff truncation of the Fock space. For
the one-mode truncation, we found a critical  coupling consistent with the best
equal-time calculations.  Increasing the Tamm--Dancoff type truncation, we
found rapid convergence with the total number of particles $N$. This is to be
contrasted with the equal-time approach where an infinite number of particles
are required to produce a critical point.

We found that above the critical coupling the zero mode develops
a contribution that breaks the $Z_2$ symmetry of the theory.
There are two such solutions to the constraint equation:
one with $\langle 0 |\phi|0\rangle >0$ and
one with $\langle 0|\phi|0\rangle <0$.
This, in turn, gives rise to two Hamiltonians
which have equivalent spectra.
The spectrum has the expected behavior:
the Fock vacuum is the state of lowest energy and the
lowest eigenvalue has a minimum at the critical coupling.
Closer inspection of the field shows that
tunneling occurs between positive and negative
eigenvalues in the broken phase.

Our renormalization prescription removes tadpoles from ordinary
interaction terms and
would properly renormalize the theory if the zero mode were removed.
However, inclusion of the zero mode appears to produces
divergences in the constraint equation and in
the resulting Hamiltonian. As a consequence, $g_{\rm critical}$
appears to grow logarithmically with the number of modes.

For $\left(\phi^4\right)_{1+1}$ we believe that a
vanishing of the mass gap is associated with the critical
coupling~\cite{hari}. We have see that the gap between
the vacuum and the lowest energy excited state
is minimized at the critical coupling.
One could imagine that, in the limit of
large number of modes, the gap between the vacuum and the first
excited state goes to zero at the critical coupling.
\section{Perturbation Theory}  \label{PertTheory}
\setcounter{equation}{0}

Even in weak coupling one must include the zero mode to correctly formulate a
light-front quantized field theory. In this phase the zero-mode constraint
equation can be solved perturbatively, and the theory that results when one
substitutes the perturbative solution for the zero mode back into the
Hamiltonian is very complicated. Nevertheless, one can systematically solve
this theory order by order in perturbation theory.

There are two important questions that one must ask about the solutions: is the
theory finite and does it agree with equal-time perturbation theory? Recall
that in a pure $\phi^4$ theory without the zero mode, the only infinities are
tadpoles, and these have been removed with a mass counterterm (via normal
ordering). Including the zero mode in the Hamiltonian introduces new
interactions that could lead to new infinities.
%
%
Comparison of perturbative results with those of
equal-time perturbation theory must be done because this is the only way we
have of checking that the zero modes have not generated a completely different
theory.

To address the renormalization problem we consider the Hamiltonian given in
Eq.~(\ref{hamiltonian}) and divide it into two parts
\begin{equation}
H= \left(g \Sigma_2+6 \Sigma_4 - C\right) +H_{\rm zm}\;.
\end{equation}
The first part contains no zero modes;
it is normal ordered and therefore totally finite.
$H_{\rm zm}$ contains all of the zero-mode interactions
\begin{eqnarray}  \label{H2}
H_{\rm zm} &=&-{a_0^4\over 4}- \frac{1}{4}\sum_{n\neq 0} \frac{1}{|n|}
\left( a_n a_0^2 a_{-n} - a_0 a_n a_{-n} a_0\right) \nonumber \\
 & & +\frac{1}{4} \sum_{k, l, m \neq 0} \frac{\delta_{k+l+m, 0}}
{\sqrt{|k l m |}}
\left( a_k a_0 a_l a_m + a_k a_l a_0 a_m\right)\; .
\end{eqnarray}
The perturbative solution of the constraint equation
(\ref{constraint}) in $1/g$ is
\begin{equation}  \label{expansion}
a_0 = -\frac{6}{g} \Sigma_3 +  \frac{6}{g^2}\sum_{n\neq 0} {1\over{|n|}}
\left( \Sigma_3 a_n a_{-n} + a_n a_{-n} \Sigma_3 + a_n \Sigma_3 a_{-n} -
\frac{3 \Sigma_3}{2} \right)+ \ldots \; . \label{pconstraint}
\end{equation}
We want to argue that $H_{\rm zm}$, with $a_0$ given by
Eq.~(\ref{pconstraint}), is finite. In two dimensions the only divergence comes
from the self-contraction  of a line: a tadpole. Notice that $H_{\rm zm}$ is
made of the product of momentum conserving interactions. It contains a two
point interaction of the form $a_n a_{-n}/|n|$, a three point interaction of
the form $ a_k a_l a_m \delta_{k+l+m,0}/{\sqrt{|k l m |}}$, and the
interactions in $a_0$. The operator $a_0$ itself contains the same two and
three point interactions. We can then consider a general process in
perturbation theory using these vertices.

The self-contraction of two lines associated with a three particle vertex would
leave the third leg of the vertex with zero momentum. Since all legs are
required to have a non-zero momentum we conclude that the three particle
vertices cannot produce tadpoles. The explicit two-particle interaction in
$H_{\rm zm}$ does give a divergence of the form
\begin{equation}
{a_0}^2\sum_{n\neq 0} {1\over{|n|}}\,.
\end{equation}
However there are two such interactions in $H_{\rm zm}$ and they cancel. The
only  remaining source of divergences are in the perturbative expansion of
$a_0$.  We see terms in the second order expansion of $a_0$ that have divergent
tadpoles of the form
\begin{equation}
\Sigma_3\sum_{n\neq 0} {1\over{|n|}}\,,
\end{equation}
but again they cancel among themselves. This story is repeated order by
order in perturbation theory, leaving a totally finite result to all
orders.

Let us take a closer look at the unbroken-phase zero mode in  perturbation
theory.  We know that the broken-phase zero mode produces spontaneous
symmetry breaking in the Hamiltonian.   What is the effect of the
unbroken phase zero mode? For the unbroken phase, the contribution to
loop integrals of the zero mode should vanish in the  infinite volume
limit, giving a ``measure zero'' contribution.
For a box of finite volume $d$, the zero mode does contribute,
compensating for the fact that the longest wavelength mode has been
removed from the system.
Thus, inclusion of the zero mode improves convergence to the infinite
volume limit~\cite{Wivoda}; it acts as a form of
infrared renormalization.
%
%
\begin{figure}
\centering\inputepsf{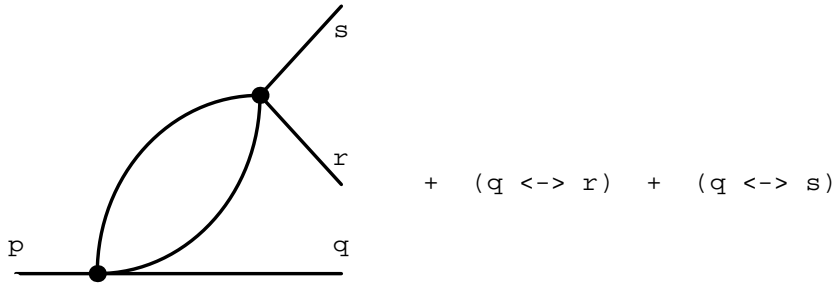}
\caption{\label{f1a}The second order contribution to the
1 particle to 3 particle matrix element [$1\to 3$].}
\end{figure}

In addition, one can use the perturbative expansion of the zero mode to
study the operator ordering problem.  One
can directly compare our operator ordering ansatz
with a truly Weyl-ordered Hamiltonian and with Maeno's operator
ordering prescription \cite{maeno}.

\begin{figure}
\centering\inputepsf{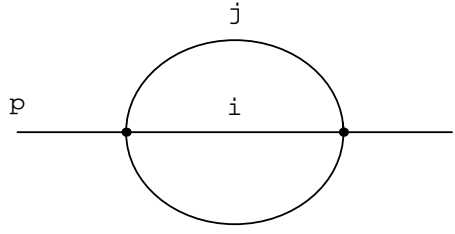}
\caption{\label{f1b}The second order contribution to
the propogator [$1 \to 1$].}
\end{figure}

As an example, let us examine $O(\lambda^2)$ contributions to the
processes $1 \to 3$ in Fig.~\ref{f1a} and $1\to 1$ in Fig.~\ref{f1b}.
First, we calculate the ``no zero
mode'' contributions.  For $1 \to 3$, we have
\begin{eqnarray}
\label{loop}
\lefteqn{\langle 0 | a_p \left(6 \Sigma_4 \right) \frac{1}{E-g \Sigma_2}
\left(6 \Sigma_4 \right)  a_q^\dagger a_r^\dagger a_s^\dagger | 0
\rangle =} \nonumber \\
& & \frac{18 \delta_{p,q+r+s}}{g \sqrt{p q r s}} \sum_{k=1}^{r+s-1}
\frac{1}{k \left(r+s-k\right)} \;\; \frac{1}{\frac{E}{g} - \frac{1}{q} -
\frac{1}{k} - \frac{1}{r+s-k}}+\nonumber \\
 & &(q \leftrightarrow r)+(q \leftrightarrow s)\,,
\end{eqnarray}
and for $1 \to 1$, we have
\begin{eqnarray}
\label{sunset}
\lefteqn{ \langle 0 | a_p \left(6 \Sigma_4 \right) \frac{1}{E-g
\Sigma_2} \left(6 \Sigma_4 \right)  a_p^\dagger | 0 \rangle =} \nonumber
\\
& & \frac{6}{g p} \sum_{i,j>0}^{i+j<p} \frac{1}{i j \left(p-i-j\right)}
\;\; \frac{1}{\frac{E}{g} - \frac{1}{i} - \frac{1}{j} - \frac{1}{p-i-j}}\,,
\end{eqnarray}
where $E$ is the unperturbed energy.
%
%
In the large volume limit $d \to \infty$, for fixed external momenta (in
terms of $P^+$) and $E \propto 1/d$ (recall that $H$ is a rescaled
Hamiltonian), the sums
can be converted to integrals.
For $1 \to 3$, one obtains
\begin{equation}
-\frac{18 \delta_{p,q+r+s}}{g \sqrt{p q r s}} \frac{4 \arctan
\sqrt{\frac{\varepsilon}{4-\varepsilon}}}{\sqrt{\left(4-
\varepsilon\right) \varepsilon}}+
(q \leftrightarrow r)+(q \leftrightarrow s)\; ,
\end{equation}
with
\begin{equation}
\varepsilon =  \left(\frac{E}{g} - \frac{1}{q}\right)\left(r+s\right) \;
,
\end{equation}
where we restrict $\varepsilon < 4$.
Likewise, for $1 \to 1$, one finds the large volume limit
\begin{equation}
- \frac{6}{g p} \frac{\pi^2}{4}\,,
\end{equation}
where we have chosen $E=g/p$. It is straightforward to show that
equal-time perturbation theory diagrams give the same result.
If one computes the $k^-$ integral
in the equal-time formulation of the theory one finds a direct connection
between the diagrams obtained in equal-time quantization and those found in
light-front quantized field theory.
This connection is particularly simple for scalar field theory in
two dimensions~\cite{ligterink}.

Now we calculate the associated zero-mode contributions.
%
%
We work to leading order in $1/g$.
Substitution of the perturbative expansion (\ref{expansion}) for $a_0$
in the Hamiltonian (\ref{H2}) produces an $O\!\left(g^{-1}\right)$ term
\begin{eqnarray}
\label{hzm}
H_{\rm zm} &=& - \frac{6}{4 g} \sum_{k,l,m \neq 0}
\frac{\delta_{k+l+m,0}}{\sqrt{|k l m|}}\left(a_k \Sigma_3 a_l a_m + a_k
a_l \Sigma_3 a_m\right) \\
\label{hzm2}
      &=&  -\frac{18 \Sigma_3 \Sigma_3}{g}-\frac{12}{g} \sum_{k,l >0}
\frac{a_l^\dagger a_l}{k l \left(k+l\right)}
+\frac{6}{g} \sum_{k,l >0} \frac{a_{k+l}^\dagger a_{k+l}}{k l
\left(k+l\right)} \nonumber \\
& &- \frac{3}{g} \sum_{k,l >0} \frac{1}{k l \left(k+l\right)} + O\!
\left( g^{-2} \right)  \; .
\end{eqnarray}
Thus, the contribution of the zero mode to the two processes is
\begin{equation}
\label{eq58}
\langle 0 | a_p H_{\rm zm} a_q^\dagger a_r^\dagger a_s^\dagger |0\rangle
= - \frac{18 \delta_{p,q+r+s}}{g \sqrt{p q r s}} \frac{1}{r+s} + \left(q
\leftrightarrow r \right)+\left(q \leftrightarrow s \right)
\end{equation}
for $1 \to 3$ and
\begin{equation}
\label{eq59}
\langle 0 | a_p H_{\rm zm} a_p^\dagger |0\rangle = - \frac{9}{g p}
\sum_{k=1}^{p-1} \frac{1}{k \left(p-k\right)} -\frac{12}{g p} \sum_{k
>0} \frac{1}{k \left(k+p\right)}
+\frac{6}{g p} \sum_{k=1}^{p-1} \frac{1}{k \left(p-k\right)}
\end{equation}
%
%
for $1 \to 1$.  We have discarded the constant term.

How does the zero mode affect convergence to the $d \to \infty$ limit?
Look at Eq.~(\ref{loop}). One can take the zero-momentum limit for each
of the internal lines; that is, the limit $k \to 0$ plus the limit $k \to
r+s$.  The result is exactly {\em twice} Eq.~(\ref{eq58}).  Likewise,
for Eq.~(\ref{sunset}), one can take limits $i \to 0$ plus $j \to 0$
plus $i+j \to p$.  The result is
\begin{equation}
- \frac{18}{g p} \sum_{k=1}^{p-1} \frac{1}{k \left(p-k\right)} \; ,
\end{equation}
exactly {\em twice} the first term in Eq.~(\ref{eq59}).  (Note that the
remaining terms of Eq.~(\ref{eq59}) are due to our particular operator
ordering ansatz.)
The appearance of this factor of two was also noted by Maeno
\cite{maeno}.
%
%
\begin{figure}
\centering\inputepsf{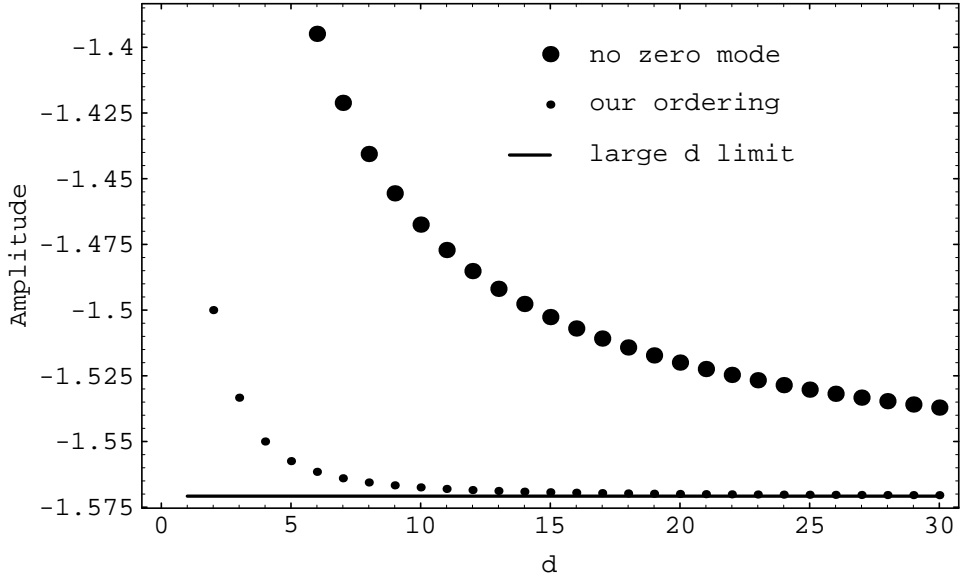}
\caption{\label{f2}Convergence to the large $d$ limit of
              $1 \to 3$ with $\varepsilon=2$.}
\end{figure}
%
%
\begin{figure}
\centering\inputepsf{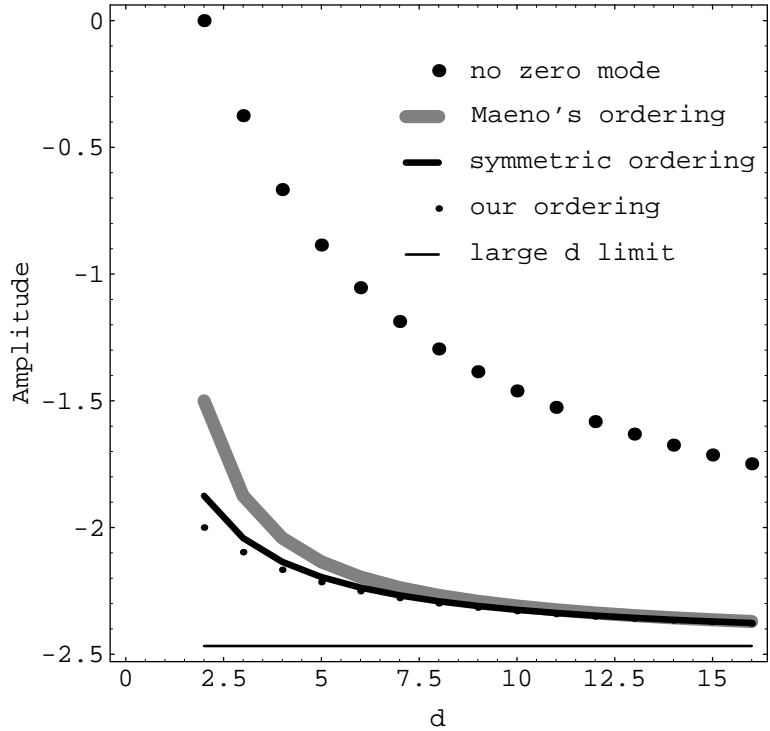}
\caption{\label{f3}Convergence to the large $d$ limit of
$1 \to 1$ setting $E=g/p$ and dropping any constant terms.}
\end{figure}
As shown in Figs.~\ref{f2} and \ref{f3}, the zero mode greatly
improves convergence to the large volume limit.
The zero mode compensates, in an optimal manner, for the fact that we
have removed the longest wavelength mode from the system.

One characteristic of our operator ordering ansatz is that the
Hamiltonian, written in terms of the dynamical variables, is no longer
symmetric ordered.  That is, Eq.~(\ref{hzm}) is not symmetric ordered.
However, in the context of perturbation theory, we are free to impose
symmetric ordering on the Hamiltonian, order by order in $1/g$.
Symmetric ordering Eq.~(\ref{hzm}) gives
\begin{eqnarray}
 -\frac{18 \Sigma_3 \Sigma_3}{g}
-\frac{9}{g} \sum_{k,l >0} \frac{a_l^\dagger a_l}{k l \left(k+l\right)}
+\frac{9}{2 g} \sum_{k,l >0} \frac{a_{k+l}^\dagger a_{k+l}}{k l
\left(k+l\right)} \nonumber \\
- \frac{9}{4 g} \sum_{k,l >0} \frac{1}{k l \left(k+l\right)} + O\!
\left( g^{-2} \right)  \; .
\end{eqnarray}
Comparing with Eq.~(\ref{hzm2}), we see that our operator ordering
ansatz and true symmetric ordering produce the same basic operator
structure with somewhat different coefficients.  Finally, applying
Maeno's operator ordering prescription to Eq.~(\ref{hzm}) produces simply
\begin{equation}
- \frac{18}{g} \Sigma_3 \Sigma_3 \; .
\end{equation}
The relative effect of the three operator orderings
on the $1 \to 1$ diagram is illustrated in Fig.~\ref{f3}.
The difference between the different orderings
is slight and decreases quickly with increasing $d$.

One can also look for the effect of the perturbative zero modes at tree
level. The simplest process where this occurs is in the second order
contributions to the six-point amplitude. There is a physical region
where zero momentum is transferred in the line connecting the two vertices.
If one omits the zero mode from this propagator the contribution of the
diagram will be zero.  Adding the zero mode leads to a more appealing
non-zero result.  If we were to compare the calculation of such a six point
amplitude with experiment, we would get conflicting results since our
zero momentum bin would give zero and the experiment would find some non-zero
result. To avoid this apparent conflict one must use
theoretical binning that is much  finer than the experimental resolution. Then
the null zero-mode contribution to lowest momentum experimental bin is small
and the issue becomes one of convergence rates, which we have already
addressed.

We can also consider making a perturbation expansion in the region where $g$
is
negative. Following Robertson~\cite{robertson} one can construct an expansion
in powers of $\sqrt{-g}$
\begin{equation}
a_0 = C_{-1} \sqrt{-g}+C_0+\frac{C_1}{\sqrt{-g}}+\frac{C_2}{\left(-g\right)}+
\frac{C_3}{\left(-g\right)^{3/2}}+ \ldots \; .
\end{equation}
We can substitute this expansion into the constraint equation and
solve order by
order for the operator coefficients $C_n$. In addition to the
unbroken-phase solution we find:
\begin{eqnarray}
C_{-1} &=& 1 \; {\rm or}\;  -1\\
C_0 &=& 0\\
C_1 &=& -3 C_{-1}\Sigma_2\\
C_2 &=& -3 \Sigma_3\\
C_3 &=& {3\over{2}} C_{-1} \sum_{n=1}^{\infty} {1\over{n}}\left( a_n \Sigma_2
a_n^\dagger +a_n^\dagger \Sigma_2 a_n -\Sigma_2 \right) - {15\over{2}}
C_{-1} \left(\Sigma_2 \right)^2\\
C_4 &=& {-21\over{2}}\left(\Sigma_2\Sigma_3 +\Sigma_3\Sigma_2\right)
+3\sum_{n=1}^{\infty}{1\over{n}} \left( a_n\Sigma_3 a_n^\dagger
+a_n^\dagger\Sigma_3 a_n -\Sigma_3\right)
\end{eqnarray}
The resulting Hamiltonian can be calculated perturbatively by inserting this
expansion for $a_0$ in Eq.~(\ref{hamiltonian}). We see that $C_{-1}$, $C_1$,
and
$C_3$ are even functions and therefore break the $Z_2$ symmetry. If we take the
vacuum expectation value of $a_0$, we find
\begin{equation}
\langle 0|a_0|0\rangle = C_{-1}\left( \sqrt{-g} +\frac{3}{2}
\zeta (2) \left(-g\right)^{-3/2}+\ldots \right)\;.
\label{pertvev}
\end{equation}
We see that for large negative $g$ (small coupling) the VEV grows like
$\sqrt{-g}$.  The expansion breaks down at $g = 0$ and is
not valid near the critical coupling. The
second term in Eq.~(\ref{pertvev}) is operator ordering sensitive and
comes from the
term  $a_n \Sigma_2 a_n^\dagger$ in $C_3$.  The Robertson calculation, for
example, does not contain this second term. Finally we see that the expansion
of
$a_0$ contains only two and three point interactions as does the perturbative
expansion in the symmetric phase. As we have argued for the symmetric
phase, the perturbative expansion is tadpole free to all orders.
\section{Nonperturbative results}  \label{NonPert}
\setcounter{equation}{0}
In I, we have shown analytically that if we truncate the problem to
one mode, the
zero mode constraint equation gives rise to critical behavior~\cite{BPV}.
Subsequent numerical studies of the one-mode and the multi-mode problem
in II have shown similar results.   The term in the
constraint equation that gives rise to spontaneous symmetry breaking is
\begin{equation}
\label{ssb}
\sum_{n > 0} \frac{1}{n}\left( a_n a_0 a_n^\dagger+a_n^\dagger a_0 a_n \right)
\; .
\end{equation}
If we take matrix elements of this term we find that it couples the
diagonal matrix elements of $a_0$ to all of the modes in the problem.
This
coupling of all of the length scales of the theory is the essential
ingredient that generates spontaneous symmetry breaking.

This gives considerable insight into how spontaneous symmetry breaking will
develop in QCD. For an operator to couple all length scales, it must be at
least third order, and it must have the structure where there is a creation
operator on one side  of the zero-mode operator and a destruction operator on
the other side. We conclude therefore that the fourth-order  interaction in QCD
is likely to be very important for spontaneous symmetry breaking.

In our previous work, I and II, we imposed a Tamm-Dancoff type
truncation on the Fock space where we limited the number of modes
$M$ and the total number of allowed particles $N$.  This allowed us
to look at the various
limits where the problem simplified.  In this paper, we will introduce a
cutoff in total momentum~\cite{PauliBrodsky}.  This cutoff is more
consistent with the fact that the zero mode $a_0$ is block diagonal
in states of equal $P^+$.
This can also be considered a cutoff in the integer-valued ``resolution''
$K_{\rm max}$ with which we probe the theory~\cite{PauliBrodsky}.
We keep all
states of the total longitudinal momentum $P^+$ such that
\begin{equation}
\frac{d}{2\pi}P^+ \le K_{\rm max}\,.
\end{equation}
As we go on to investigate this problem with finer and finer resolution we are
able to analyze a number of issues that are unique to the broken phase of the
problem near the critical coupling.

We have argued in the previous section that the standard mass subtraction,
which
has the form
\begin{equation}
\label{massct}
           -a_0 \sum_{n> 0} {1\over{n}}\,,
\end{equation}
removes the divergence in Eq.~(\ref{ssb}) and gives a totally finite theory in
perturbation theory, even for the broken phase. In the nonperturbative solution
we do not calculate order by order as in perturbation theory, and such a
cancellation is not assured.   We already saw indications in II that the
critical coupling was growing with the number of modes in problem. In the
numerical approach we use here this translates into a logarithmic divergence
associated with the $K_{\rm max}$ cutoff.

For a finite resolution $K_{\rm max}$ the constraint equation
(\ref{constraint}) becomes a finite set of coupled cubic equations.
These equations can then be solved numerically;
a simple iterative method related to
Newton's method was used to produce
the results discussed here.
To construct the iteration formulas, we
write the matrix elements of the right-hand side of (\ref{constraint}) as
functions of $g$ and the matrix elements of $a_0$
\begin{eqnarray} \label{Falphabeta}
F_{\alpha, \beta}(g,a_0)&=&\langle \alpha |\left\{
g a_0 + a_0^3 + 2 a_0 \Sigma_2 + 2 \Sigma_2 a_0 \right. \nonumber\\
& &+ \sum_{n > 0} {1\over n}
\left(  a_n^\dagger a_0 a_n + a_n a_0 a_n^\dagger - a_0 \right)+
\left. 6 \Sigma_3 \right\}|\beta\rangle \;,
\end{eqnarray}
where $|\alpha\rangle$ and $|\beta\rangle$ are generic Fock states allowed
by the $K_{\rm max}$ cutoff.  The constraint equation is then
expressed as a set of algebraic equations
\begin{equation} \label{AlgebraicConstraint}
F_{\alpha, \beta}(g,a_{0})=0
\end{equation}
that determine all the matrix elements of $a_0$ as functions of $g$.
However, in the following, we will treat
$\langle 0 | a_0 |0\rangle$ as the independent variable and $g$ as one
of the dependent variables.

The nonlinear system (\ref{AlgebraicConstraint}) can be solved
iteratively according to the following scheme:
\begin{eqnarray} \label{iteration}
g^{(n+1)}&=&
    g^{(n)}-\omega \frac{F_{0,0}(g^{(n)},a_0^{(n)})}
                    {\langle 0 |a_0|0\rangle}\;,  \\
%
%
%
%
\langle \alpha |a_0^{(n+1)}|\beta\rangle &=&
    \langle \alpha|a_0^{(n)}|\beta\rangle
      -\omega \frac{F_{\alpha, \beta}(g^{(n)},a_0^{(n)})}
                 {\left.\frac{ \partial F_{\alpha, \beta}(g^{(n)},a_0)}
            {\partial \langle \alpha
|a_0|\beta\rangle}\right|_{a_0=a_0^{(n)}}}\;,
\end{eqnarray}
where neither $|\alpha\rangle$ nor $|\beta\rangle$ is the vacuum state.
The superscript $(n)$ indicates the iteration number, and
$\omega$ is an underrelaxation parameter that should usually be chosen
to be between $0.5$ and $0.75$. This approach works well for values of
$\langle 0 |a_0|0\rangle$ much less than one and values
of $K_{\rm max}$ less than 20, but otherwise converges slowly.

As part of the truncation process, assumptions must be made about the behavior
of the zero-mode matrix elements  for large momentum $P^+$ sectors.  This is
because each matrix element is coupled to an infinite sequence of other matrix
elements through the sum that appears in the constraint
equation~(\ref{constraint}). If one assumes that matrix elements decrease
quickly to zero with increasing $P^+$  then the sum can be truncated at
$n=K_{\rm max}$; the remainder is  guaranteed to be a small correction.  This
is what we have assumed in I and II.

%
Let us examine the asymptotic behavior of the zero-mode matrix elements
more closely.
We might expect that, in the limit of large $n$, the terms
$\langle \alpha|
(a_n^\dagger a_0 a_n+a_n a_0 a_n^\dagger )|\beta\rangle/n$
will cancel the term $-\langle \alpha| a_0 |\beta\rangle/n$
in the sum over $n$ in the constraint equation~(\ref{constraint}).
After all, this is why there are no tadpoles in perturbation theory.
For the moment let us use this as our assumption for the Fock
space truncation.
%
%
For example, we will demand that the single-particle matrix
element $\langle 0| a_K a_0 a_K^\dagger |0\rangle$ be
equal to $\langle 0|a_0|0\rangle$ for $K>K_{\rm max}$.
%
%
\begin{figure}
\centering\inputepsf{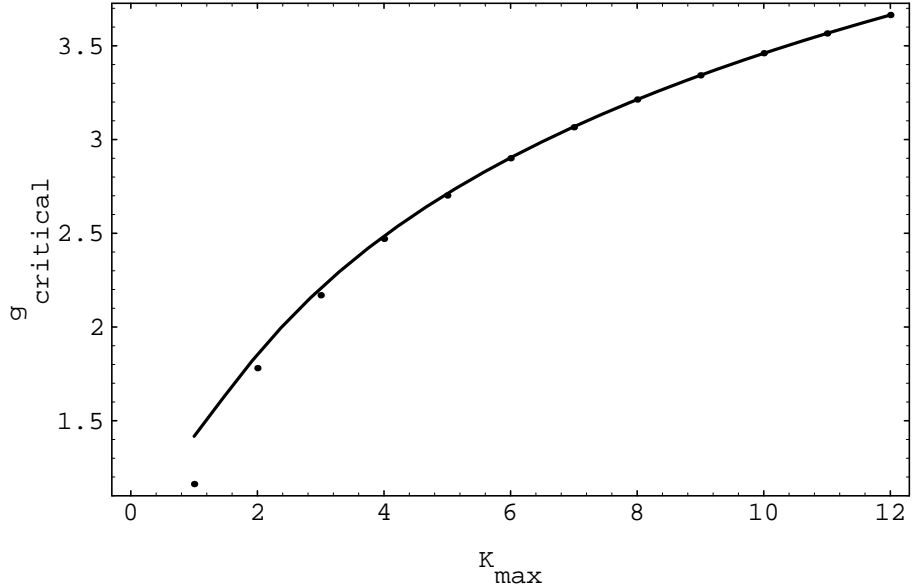}
\caption{\label{f6}Plot of $g_{\rm critical}$ vs
$K_{\rm max}$   The solid line is the
fit given in Eq.~(\protect\ref{fitforNonZeroAsymp}) of the text.}
\end{figure}
Unfortunately, if we examine the resulting  numerical solutions, we find that
the zero-mode matrix elements are inconsistent with this assumption
and that there is a
logarithmic divergence in $g_{\rm critical}$, which is fit by the
following form:
%
%
\begin{equation}  \label{fitforNonZeroAsymp}
g_{\rm critical}= 1.192 \ln K_{\rm max} + 0.638 + \frac{0.779}{K_{\rm max}}
%
%
\;.
\end{equation}
The values obtained are plotted in Fig.~\ref{f6}.
%
%
\begin{figure}
\centering\inputepsf{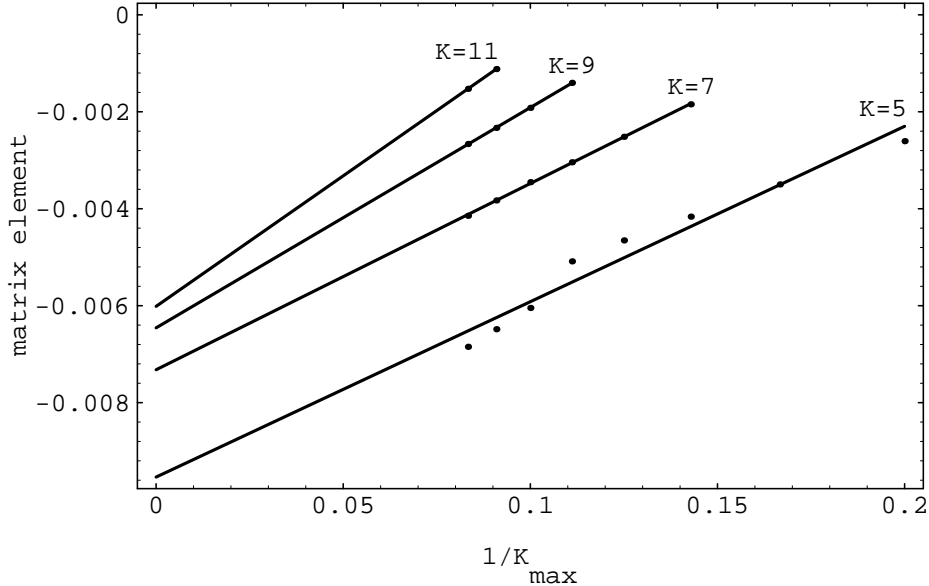}
\caption{\label{f7}Single-particle matrix elements
$\langle 0| a_K a_0 a_K^\dagger |0\rangle$
plotted for a series of values of the maximum resolution $K_{\rm max}$.
The values of $K$
selected are 5, 7, 9, and 11.  The vacuum expectation value is
set at 0.05. }
\end{figure}
This divergence
appears to be due to the fact that the mass counterterm
Eq.~(\ref{massct}) does not cancel the divergence in Eq.~(\ref{ssb}).
This behavior is illustrated in
Fig.~\ref{f7} for a calculation where
$\langle 0|a_0|0\rangle=0.05$; the asymptotic value
of $\langle 0| a_K a_0 a_K^\dagger |0\rangle$ is
between -0.003 to -0.006, which is certainly inconsistent with
$\langle 0|a_0|0\rangle=0.05$.
The difference between
the assumed and actual behaviors produces the logarithmic dependence
on $K_{\rm max}$.

%
%
\begin{figure}
\centering\inputepsf{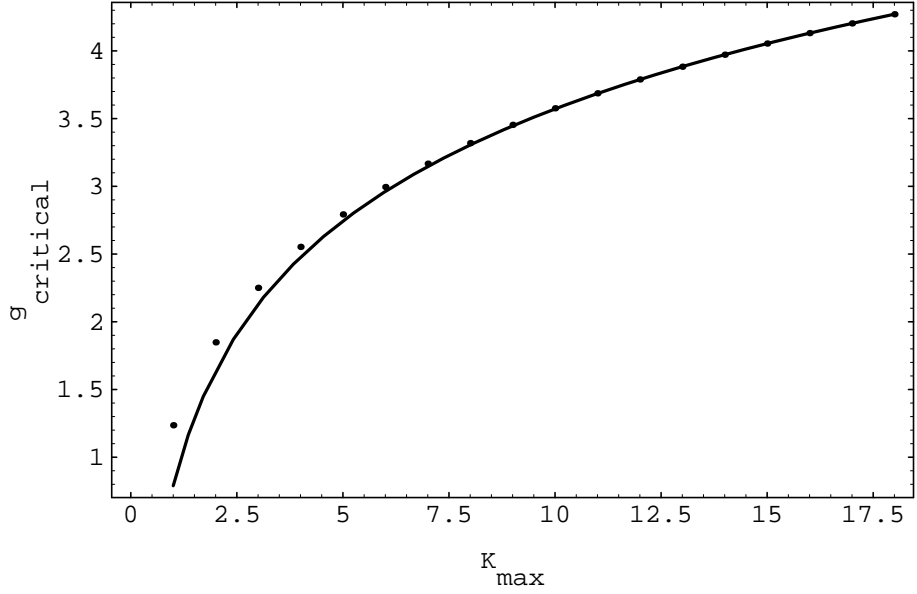}
\caption{\label{f4}Plot of $g_{\rm critical}$ vs $K_{\rm max}$ for
the case where matrix elements are
assumed to be asymptotically zero for large momentum.  The
solid line is the fit given in Eq.~(\protect\ref{fitforZeroAsymp}) of the
text.}
\end{figure}
If the truncation of the system of equations is done with the
assumption that the matrix elements tend to zero (as we did in I and II),
then different results
are obtained.  The term $-a_0 \sum_n 1/n$
is immediately logarithmically divergent and guarantees a
logarithmically divergent behavior for $g_{\rm critical}$.
However, the result for $g_{\rm critical}$, shown in Fig.~\ref{f4}, has
an additional logarithmic divergence.  It is well fit by the form
%
%
%
\begin{equation} \label{fitforZeroAsymp}
g_{\rm critical}= \sum_{n=1}^{K_{\rm max}}\frac{1}{n}+
             0.214 \ln K_{\rm max} + 0.147
          + \frac{0.0920}{K_{\rm max}}
\;.
\end{equation}
This is consistent with the results found in II and implies
that the broken phase of the theory  requires additional
renormalization.  Clearly, however, one cannot remove the logarithmic
divergence in $g$ by simply performing a different subtraction that
would eliminate the $-a_0 \sum_n 1/n$ term in the constraint equation.
%
%
\begin{figure}
\centering\inputepsf{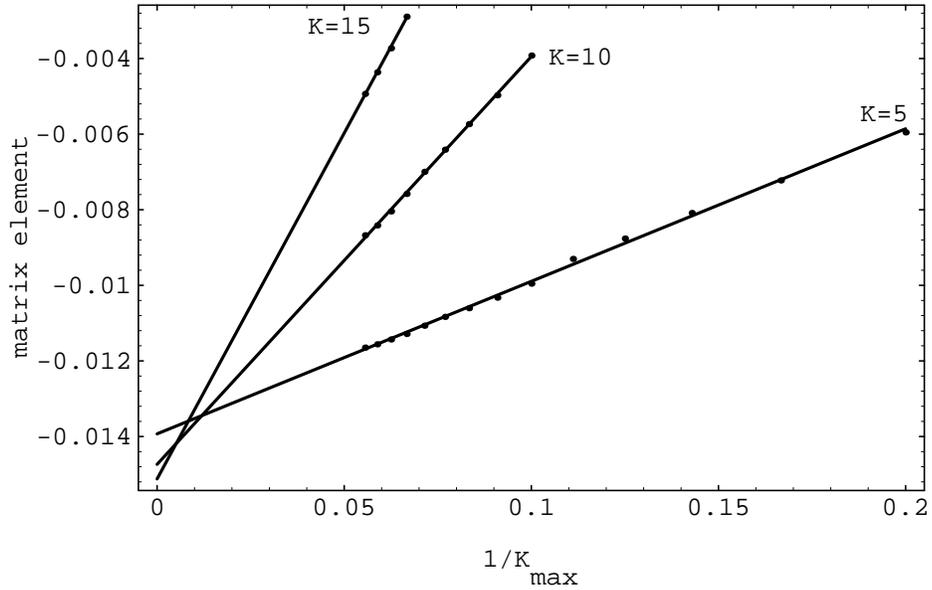}
\caption{\label{f5}Single-particle matrix elements
$\langle 0| a_K a_0 a_K^\dagger |0\rangle$
plotted for a series of values of
the maximum resolution $K_{\rm max}$.
The values of $K$
selected are 5, 10, and 15.  The vacuum expectation value is
set at 0.05.  The matrix elements are calculated under the
assumption that they are asymptotically zero.}
\end{figure}
The extra divergence comes from
the behavior of matrix elements of $a_0$ in the limit of
large momenta.  Instead of going to zero, they tend toward a nonzero constant.
As an illustration of this,  one-particle matrix elements are
shown in Fig.~\ref{f5}.  This behavior is inconsistent with the original
assumption that the matrix elements go to zero in the limit of large momentum.

\begin{figure}
\centering\inputepsf{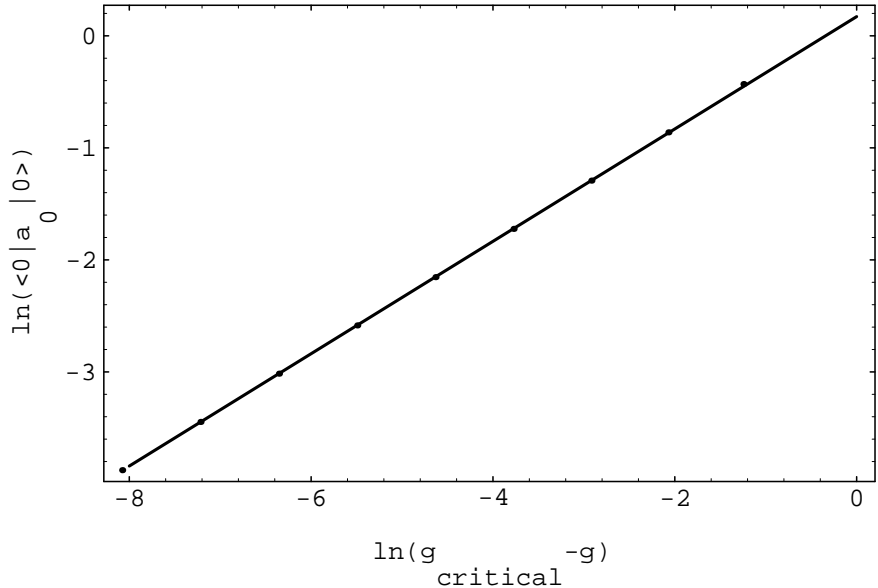}
\caption{\label{f8}Plot of $\ln \left(\langle 0 | a_0 | 0 \rangle\right)$ vs
$\ln \left(g_{\rm critical}-g\right)$, with $K_{\rm max}=10$.  The slope yields
the critical exponent $\beta$, which appears in
Eq.~(\protect\ref{gamma}) of the text.}
\end{figure}

We can also investigate the shape of the critical curve near the critical
coupling as we increase the resolution. The general theory of critical
behavior indicates that the VEV should behave as a power law near the
critical coupling~\cite{huang}
\begin{equation}  \label{gamma}
{\rm VEV}\propto(g_{\rm critical}-g)^\beta\; .
\end{equation}
We have calculated the exponent $\beta$ as a function of the resolution,
$K_{\rm max}=5$ to 12, and find it to be equal to $0.50$, independent of
$K_{\rm max}$. A typical fit, where $\beta$
is the slope, is shown in Fig.~\ref{f8}.
However, $\left(\phi^4\right)_{1+1}$
is in the same
universality class of theories as the Ising model, and for this model
it is known that the exact value of $\beta$ is $1/8$. Of course we would
have been surprised if we obtained the exact
answer near the critical point using an approximation method as
primitive as DLCQ.
Because we have studied the long range behavior of this
theory with one mode of zero momentum, we should reasonably expect to
obtain results that are about as good as those of a mean field
calculation in the equal-time approach
and, indeed, the value of the critical exponent that one obtains
for the Ising model in the mean field approximation is $1/2$.

\section{Conclusion} \label{Conclusion}

Our analysis of this rather simple two-dimensional scalar theory has produced a
number of very interesting and important results for the program of solving QCD
as a light-cone quantized field theory. The most important is that we can
understand a variety of long-range phenomena starting from a simple Fock-space
vacuum. We have found that the long-range physics of this theory is uniquely
contained in a set of additional operators.  In the DLCQ formulation these
operators are determined, up to renormalizations, by an operator constraint
equation (\ref{constraint}). Furthermore, the theory that contains the new
operators agrees with equal-time perturbation theory. Thus one is confident in
this formulation that we are solving the theory that we set out to solve.

There are a number of ways in which this analysis is incomplete, and the
reasons for this are important for our QCD program~\cite{kalloniatis}.
One is that we have not
succeeded in renormalizing the new operators in the broken phase of the theory.
The reason is that in two dimensions there are an infinite number of allowed
operators. Thus the new operators introduced through the zero modes can be very
complicated, and in fact {\em are} very complicated, making renormalization
difficult.  Unfortunately it has been shown that the number of allowed
operators in QCD can be very large as well, and therefore we might find similar
difficulties there.

Another way in which the analysis is incomplete is that the solutions
to the constraint equation are not completely consistent with our
Fock space truncation.  This behavior is related to the
logarithmic divergence of $g_{\rm critical}$ and is not well understood.
It is unclear if similar problems will arise in a study of QCD.

A third deficiency is evident from the calculation of the critical exponent
$\beta$.  A complete analysis should yield $\beta=1/8$ while we have obtained
the mean-field value of $1/2$. Typically a mean field result indicates that one
has not included enough length scales in the calculation.  We used DLCQ as a
regulator and also as a calculational technique: we chose our momenta on a
linear scale in the  numerical calculations.  Thus, we were not able to include
many length scales in the numerical calculations. We need to find a way of
improving the numerical approach so that many length scales are included.

\section*{Acknowledgments}
The authors would like to thank the Stanford Linear Accelerator Center,
where much of this work was completed, for its hospitality and support.
The authors would also like to acknowledge Stan Brodsky and
Marvin Weinstein for many useful comments. This work
was supported in part by grants from the U. S. Department of Energy
and in part by grants of computer time from the Minnesota
Supercomputer Institute and the support of a NATO collaborative grant.


\begin{thebibliography}{99}

%
%









%
%








\bibitem{BPV}
C. M. Bender, S. S. Pinsky, and B. van de Sande,
Phys.\ Rev.\ D {\bf 48}, 816 (1993).

\bibitem{PV}
S. S. Pinsky, and B. van de Sande,
Phys.\ Rev.\ D {\bf 49}, 2001 (1994).

\bibitem{maskawa} T. Maskawa, and K. Yamawaki, Prog.\ Theor.\ Phys.\
{\bf 56}, 270 (1976).

\bibitem{heinzl} T. Heinzl, S. Krusche, S. Simburger, and E. Werner,
Z. Phys.\ C {\bf 56}, 415 (1992);
Heinzl, S. Krusche, and E. Werner,
Phys.\ Lett.\ B {\bf 272}, 54 (1991);
Heinzl, S. Krusche, and E. Werner,
Phys.\ Lett.\ B {\bf 275}, 410 (1992);
Heinzl, S. Krusche, and E. Werner,
Nucl.\ Phys.\ A {\bf 532}, 4290 (1991).

\bibitem{wittman}
R. S. Wittman,
in {\em Nuclear and Particle Physics on the Light
Cone}, edited by M. B. Johnson and L. S. Kisslinger
(World Scientific, Singapore, 1989).

\bibitem{PauliBrodsky}
H.-C. Pauli and S.J. Brodsky, Phys.\ Rev.\ D
{\bf 32}, 1993 (1985); {\bf 32}, 2001 (1985).

\bibitem{PW} S. J. Brodsky, G. McCartor, H. C. Pauli, and S. S. Pinsky,
Particle World  {\bf 3}, 109 (1993);
S. J. Brodsky and H. C. Pauli,
in {\em Recent Aspects of Quantum Fields,}
edited by H. Mitter and H. Gausterer, Lecture Notes in Physics Vol.\ 396
(Springer-Verlag, Berlin, 1991).
{\em Proceedings of the 30th Schladming Winter School, 1991}.

\bibitem{benderpinsky}
C. M. Bender, L.R. Mead, and S. S. Pinsky,
Phys.\ Rev.\ Lett.\ {\bf 56}, 2445 (1986).

\bibitem{hari}
A. Harindranath and J. P. Vary,
Phys.\ Rev.\ D {\bf 37}, 1076 (1988);
A. Harindranath and J. P. Vary,
Phys.\ Rev.\ D {\bf 36}, 1141 (1987).

%
%
\bibitem{Wivoda}
J.J. Wivoda and J.R. Hiller,
Phys.\ Rev.\ D {\bf 47}, 4647 (1993).

\bibitem{maeno} M. Maeno, Phys.\ Lett.\ B {\bf 320}, 83 (1994).

\bibitem{ligterink}N. E. Ligterink and B. L. G. Bakker,
``Equivalence of Light-Cone and Instant-Form Dynamics,''
unpublished.

\bibitem{robertson}
D. Robertson, Phys.\ Rev.\ D {\bf 47}, 2549 (1993).


\bibitem{huang} Kerson Huang, {\em Statistical Mechanics}, (John Wiley \& Sons,
New York, 1987).

\bibitem{kalloniatis}A. C. Kalloniatis, H. C. Pauli, and S. S. Pinsky,
Ohio State preprint No.\ \goodbreak OHSTPY-HEP-TH-94-001 (unpublished).



\end{thebibliography}
\end{document}